University of Basel

Faculty of Psychology

# Needs, Passions and Loot Boxes – Exploring Reasons for Problem Behaviour in Relation to Loot Box Engagement

Author: Cooper, Dylan Mercury

Matriculation number: 20-050-266

Correspondence email: dylan.cooper@stud.unibas.ch

# Contents

**Abstract**

Research on the convergence of gaming and gambling has been around since the 1990s. The emergence of loot boxes in video games in the mid 2010s, a game mechanic with a chance-based outcome that shares structural and psychological similarities to gambling, caused public controversy and led to the inception of a new field of study, loot box research. Since then, various studies have found a relationship between loot box engagement and problem gambling as well as problem gaming. Due to the cross-sectional nature of this data, however, inferences about causality are limited. While loot box research has extensively investigated the relationship between loot box engagement and problem behaviour, little research has been done to explain the underlying motivations of players that drive them to interact with loot boxes. The goal of this thesis is to provide possible explanations for the relationship between loot box engagement and problem gamblers or problem gamers. In doing so, it draws upon two prominent psychological theories: Self-Determination Theory and the Dualistic Model of Passion. Self-Determination Theory's concept of psychological needs and their satisfaction or frustration is hereby used to explain the development of harmonious or obsessive passions, which are introduced in the Dualistic Model of Passion. These obsessive passions have been shown to be possible antecedents of behavioural addictions, such as problem gambling or problem gaming. Thus, the interplay between needs, passions and loot box opening could elucidate the aforementioned correlations between loot box engagement and problem behaviour. However, further research, especially utilising longitudinal data, is needed to better understand these processes.

**Introduction**

Loot boxes are a novel game mechanic which contain chance-based outcomes and have been the cause of public controversy, especially in the gaming community, since 2017 (Drummond & Sauer, 2018; McCaffrey, 2022). Their form, availability and nomenclature appear with great variation, which is why their academic pursual and understanding has proven difficult so far. Generally, loot boxes are defined as consumable virtual items, which can be redeemed for a randomised selection of rewards. These rewards can range from items which have only cosmetic effects, to items which have game-changing "pay to win" features, for instance improved weapons or power ups (Gibson et al., 2022; Spicer et al., 2022).

Loot box prevalence in video games has risen significantly over the last decade, with loot boxes now being an integral part of all major game franchises (Zendle et al., 2020). Moreover, estimates show that 44-78% of players have thought about purchasing loot boxes (Li et al., 2019; Spicer et al., 2022; Zendle & Cairns, 2018). This statistic is especially concerning when considering that 95% of iPhone and 93% of Android games which contain loot boxes are deemed suitable for ages 12 and up (Advisory Board for Safer Gambling UK, 2021). Due to loot boxes being so widely available to children they have come under increasing scrutiny from the media, policymakers, academics and the gaming community itself, especially due to their structural and psychological similarities to gambling (Drummond et al., 2020; Drummond & Sauer, 2018; McCaffrey, 2022; Petrovskaya & Zendle, 2022).

This led to multiple studies being conducted, searching for a possible relationship between loot boxes and, primarily, problem gambling. Over the past years, positive correlations between loot box opening and problem behaviour, in particular problem gambling and problem gaming, have repeatedly been found (Gibson et al., 2022; Montiel et al., 2022; Spicer et al., 2022). The cross-sectional nature of this data, however, makes it challenging to draw conclusions about causality. In addition to that, the methodological quality of the majority of studies has been questioned, leading to possible problems of generalisability, sample selection bias and a neglect of one of the most important loot box demographics, children (Gibson et al., 2022; Spicer et al., 2022; Yokomitsu et al., 2021).

In addition to neglecting important demographics, another area which has received insufficient attention in the field of loot box research, is the investigation into the underlying motivations for loot box engagement. Hereby the use of two prominent theories in psychology, Self-Determination Theory (SDT) and the Dualistic Model of Passion (DMP), were drawn upon to provide potential explanations for the relationship between loot box engagement and problem gambling or problem gaming.

Self Determination Theory's mini theory, the Basic Psychological Needs Theory (BPNT), proposes three basic psychological needs; *autonomy*, *competence*, and *relatedness*, which when satisfied are associated to psychological well-being, vitality and overall contentment. The frustration of these needs, however, has been linked to psychological distress and behavioural addictions, such as problem gambling and problem gaming (Allen & Anderson, 2018; Chen et al., 2015; Costa et al., 2016; Mills et al., 2018; Mills & Allen, 2020; Vansteenkiste et al., 2020; Vansteenkiste & Ryan, 2013).

When an activity satisfies one's needs, it has potential to consequently turn into a passion. Hereby, the Dualistic Model of Passion proposes two forms of passion, *harmonious* and *obsessive* passion. Harmonious passion (HP) results from an autonomous internalisation of an activity into a person's identity and is often associated with greater life satisfaction, whereas obsessive passion (OP) results from a controlled internalisation and is associated with an urge to engage with the activity, even if it is detrimental to other, objectively more important, life domains and responsibilities (Davis & Arterberry, 2019; Mageau & Vallerand, 2007; Seguin-Levesque et al., 2003; Steers et al., 2015; Vallerand, 2015; Verner-Filion et al., 2017).

Additionally, past research has not only found that activity-specific need satisfaction can lead to obsessive passions but that activity-specific need satisfaction, in relation to global need satisfaction, can be an antecedent for problem behaviour. This means, situations in which context-specific need satisfaction is high in comparison to global need satisfaction have shown to be at risk for the development of behavioural addictions. Various studies have shown that the combination of low global need satisfaction and high context-specific need satisfaction can be an antecedent of problem gambling and problem gaming (Allen & Anderson, 2018; Bender & Gentile, 2020; Holding et al., 2021; Mills et al., 2018; Przybylski et al., 2009; Rigby & Ryan, 2011).

This thesis aims to provide possible explanations for the relationship between loot box engagement and problem gamblers or problem gamers. In doing so, literature on behavioural addictions in relation to the Dualistic Model of Passion and Self-Determination Theory are used to postulate possible links between the satisfaction or frustration of basic needs, the development of passions and symptoms of problem behaviour in respect to loot boxes.

**Methods**

The literature search for this thesis began with keywords such as "loot boxes", "problem gambling" and "problem gaming" to create a fundamental understanding of the current state of the literature. Hereby systematic, scoping and other literature reviews, such as "Loot Boxes, Problem Gambling, and Problem Gaming: A Critical Review of the Emerging Literature" by McCaffrey (2022) provided a good overview of the main findings in the literature. Other reviews such as the papers by Spicer et al. (2022), Montiel et al. (2022), Yokomitsu et al. (2021), Garea et al. (2021) and Gibson et al. (2022) and papers which had a noticeable impact on the field, such as "Video game loot boxes are linked to problem gambling: Results of a large-scale survey" by Zendle and Cairns (2018) and "Video game loot boxes are psychologically akin to gambling" by Drummond and Sauer (2018) were used to gain further insight into the field.

Additionally, in hopes of postulating one or more explanations for the relationship between problem behaviour, problem gambling in particular, and loot box engagement, two established psychological theories were utilised; the Dualistic Model of Passion by Vallerand (2015) and Self-Determination Theory by Ryan and Deci (2017). These were used to explore potential explanations for problem behaviour in relation to loot box engagement by citing already established work on the associations between basic psychological needs, passions and behavioural addictions (Allen & Anderson, 2018; Holding et al., 2021; Lalande et al., 2017; Michalczuk et al., 2011; Mills et al., 2021; Mills et al., 2018; Mills & Allen, 2020; Vuorinen et al., 2022). Supplementary literature was used to solidify points being made and gain a more in-depth understanding. To manage references the citation program "Mendeley" was used.

## Theoretical Background

**Self-Determination Theory**

Self-Determination Theory (SDT) is a macro-theory of human behaviour and personality development, whereby its main focus lies on social conditions and how these improve or hinder human flourishing. SDT demonstrates how contexts can benefit or undermine the satisfaction of needs and motivations which are needed for effective self-regulation and psychological well-being. By identifying and analysing different kinds of motivational regulation and the conditions which cultivate them, SDT can be applied, systematically, in various contexts. Hereby the adequacy of different contexts is compared to one another, in an attempt to optimise well-being. In doing so, SDT studies intrinsic and multiple extrinsic types of motivation and describes how these, along with the satisfaction or frustration of three basic psychological needs, can influence behavioural patterns, for instance those of behavioural addiction (Bonnaire et al., 2016; Botterill et al., 2016; Edgren et al., 2016; Hardoon et al., 2004; Hing et al., 2017; Kotov et al., 2010; Michalczuk et al., 2011; Mills & Allen, 2020; Nower et al., 2004; Nower & Blaszczynski, 2017; Petry et al., 2008; Ryan & Deci, 2017; Tabri et al., 2017; Vachon & Bagby, 2009).

*The Motivation Continuum*

SDT makes it clear that understanding *why* individuals do things is vital to understanding *how* they do those things. The way people typically interact with their environment is indicative of their general motivations, which is a key component of SDT. According to SDT, three inter-connected motivations are what generally distinguish individuals from one another. These three types of general motivation lie on a continuum, which ranges from strong, intrinsic motivation, to extrinsic motivation, which is split up into four different types and lastly amotivation, which is also known as a lack of motivation (Ryan & Deci, 2017).

SDT poses that individuals are primarily intrinsically motivated when they follow *autonomous* motivations which are driven by their own interest of gaining new knowledge, reaching new goals or witnessing new emotional and physical sensations (Carbonneau et al., 2012; Gillet et al., 2012; Tóth-Király et al., 2019). As opposed to autonomous motivations, individuals acting with *controlled* motivations, which are extrinsically motivated, often possess insecurities with their self-perception which results in a need to defend and demonstrate their sense of self-worth (Hodgins & Knee, 2002; Kernis, 2003; Lewis & Neighbors, 2005). This can lead them to choose activities and environments which provide them with their desired approval. Lastly and expanding upon the continuum of autonomous and controlled motivations, *amotivation* is present in people, who largely attribute outcomes to fate or factors beyond their control, due to strong feelings of inadequacy. Adaptive outcomes which are autonomously motivated have shown to be associated with life satisfaction while amotivation and controlled motivation is associated with more psychological distress (Ryan & Deci, 2000, 2017).

*Basic Psychological Needs Theory*

Well-being, as described by SDT, is not simply the absence of negative feelings. It is characterised as the ability to thrive or be fully functionable in life and is accompanied by a sense of vitality, awareness and access to, as well as exercise of one's human capacities and true regulation of the self. A proposition made by Basic Psychological Needs Theory (BPNT), a mini theory of SDT, is that the optimal development and integrity of well-being is comprised of three basic psychological needs and their satisfaction; *autonomy*, *competence* and *relatedness*. BPNT argues that these needs are innate and universal, existing across individuals and cultures alike (Legault, 2017; Ryan, 2009; Ryan & Deci, 2000, 2017). Interestingly, BPNT, especially the concepts of need satisfaction and need frustration have often been used to study video game player experience in past research (Tyack & Mekler, 2020).

**Need Satisfaction and Need Frustration**

Need satisfaction and need frustration, a further concept introduced by SDT, is a result of the relationship between individuals and their environment (Ryan & Deci, 2000, 2017; Vansteenkiste & Ryan, 2013). Need satisfaction is satiated when individuals feel fulfilled in their three basic psychological needs. Examples for need satiation in people could be; *autonomy* - Perceiving their actions to be in their control, *competence* - feeling capable of accomplishing their tasks, and *relatedness* - feeling cared for and connected to others (Ryan & Deci, 2000; Vansteenkiste & Ryan, 2013). As previously mentioned, need satisfaction is essential for psychological well-being and thriving and is positively associated with positive emotions, feelings of liveliness and even physical health (Gunnell et al., 2016; Ryan & Deci, 2017; Ryan & Frederick, 1997; Sheldon et al., 1996).

In contrast to need satisfaction, need frustration occurs when social constructs hinder people from satisfying their basic needs which in turn results in raised psychological distress (Chen et al., 2015). This could either mean a perceived lack of competence due to feeling inferior to others, lack of autonomy due to a sense of forced behaviour or reduced relatedness after experiencing rejection in some form (Chen et al., 2015; Mills et al., 2020; Mills, et al., 2018; Vansteenkiste et al., 2020).

**The Dualistic Model of Passion**

Passion is defined by Vallerand (2015) as:

A strong inclination toward a specific object, activity, concept or person that one loves (or at least strongly likes), highly values, invests time and energy in on a regular basis, and that is part of one's identity. Furthermore, two forms of passion seem to exist. The first can be seen as being in harmony with other aspects of the self and the person's life and should mainly lead to adaptive outcomes. The second form of passion may conflict with aspects of the self and the person's life and should mainly lead to less adaptive, and sometimes, even maladaptive outcomes. (p. 41)

The Dualistic Model of Passion (DMP), also posited by Vallerand (2015), makes the assumption that people are inherently inclined to internalise substantial elements of their environment into their identity, which is a crucial tendency towards people's self-growth. In line with Self-Determination Theory, DMP proposes that internalisation transcends the internal/external dichotomy and instead occurs on a continuum which can range from fully autonomous to fully controlled. Autonomous internalisation takes place when individuals, of their own volition, accept the activity as important, without any interference from the social environment. These internalisations are aligned with the integrated self and can facilitate self-growth in cognitive, regulative, affective, interpersonal, and societal dimensions. In contrast, controlled internalisation processes take place when the social environment acts in a controlling way which forces us to behave in its interest. These types of internalisations usually emanate from outside the integrated self, known as the ego-invested self, which attaches feelings of social acceptance or self-esteem to the activity. This can limit or even prevent adaptive self-processes and can lead to maladaptive outcomes. The type of internalisation process that occurs determines the resulting type of passion, of which there are two (Chandler & Connell, 1987; Deci & Ryan, 1985; Kernis, 2003; Ryan & Deci, 2000, 2017; Sheldon, 2002; Vallerand, 1997, 2015). Crucially, and relating to SDT, passions often result from activities which satisfy our needs and consequently are more likely to be valued (Vallerand, 2015).

***Harmonious and Obsessive Passion***

The Dualistic Model of Passion, just as the name implies, differentiates between two types of passions, *Harmonious* and *Obsessive* Passion. *Harmonious Passion* (HP) results from an autonomous internalisation of the activity into the person's identity and is characterized by a balanced involvement. The individual freely chooses to engage with the activity, without feeling an urge to do so and can refrain from it, if the circumstances so require. The activity is hereby in harmony with other self-elements and aspects of the individual's life, occupying a substantial but not overbearing part of their identity. They can enjoy the benefits of the activity, without compromising other goals or responsibilities, thus showing flexibility in their behavioural engagement. In contrast, *Obsessive Passion* (OP) results from a controlled internalisation of the activity into the person's identity. This results in the individual's experiencing an uncontrollable urge to engage with the activity. Even though people with an obsessive passion love, or at least like the activity, they feel compelled to engage with it, even if it is in conflict with other important goals and responsibilities (Holding et al., 2021; Mageau & Vallerand, 2007; Seguin-Levesque et al., 2003). This can, to some extent, lead to activity-specific advantages, for example performance improvement. However, these benefits generally don't overshadow the long-term costs linked to OP (Verner-Filion et al., 2017). Research has consistently shown that HP is associated with positive outcomes, such as life satisfaction, while activities moulded by OP negatively affect well-being in numerous ways (Curran et al., 2015; Lafrenière et al., 2009; Vallerand, 2015).

***Need Satisfaction and Passion Development***

In the past, research has shown that high need satisfaction in the context of an activity is positively related to both HP and OP, with the relation to HP however being stronger than the relation to OP. Additionally, investigations into the effect of need satisfaction outside the realm of an activity, for example life in general, were made. These found that general, also known as *global* need satisfaction, as opposed to the activity's *context-specific* need satisfaction, was negatively associated with OP. This suggests that individuals are less likely to score high in levels of obsessive passion if their needs in other life domains outside the activity are met. This implies that developing an OP may represent a way of compensating need satisfaction, particularly if global need satisfaction (i.e., need satisfaction outside of the activity) is low. This is in contrast to harmonious passion which showed no relation to global need satisfaction (Holding et al., 2021).

**Problem Gambling and Problem Gaming**

Problem gambling and problem gaming are both behavioural addictions, with problem gambling being defined by Ferris and Wynne (2001) as:

> Gambling behaviour that creates negative consequences for the gambler, others in his or her social network, or for the community. (p. 8)

It is predominantly measured using the Problem Gambling Severity Index (PGSI) (Ferris & Wynne, 2001; Garea et al., 2021; Holtgraves, 2009; McCaffrey, 2022).

Problem gaming or excessive gaming, on the other hand, is commonly known as either Gaming Disorder, which was added to the World Health Organization's (WHO) International Classification of Diseases (ICD-11), or Internet Gaming Disorder, which was added to the Diagnostic and Statistical Manual of Mental Disorders (DSM-5) and is measured with the accompanying symptomology scale (American Psychiatric Association, 2013; World Health Organization, 2021). Although these classifications are controversial and, especially IGD, have been heavily criticised, they are widely used in gaming research (McCaffrey, 2022; van Rooij et al., 2018; Yokomitsu et al., 2021).

*Self-Determination Theory and Behavioural Addictions*

As previously described, Self-Determination Theory can be used to understand the development of problem behaviour and behavioural addictions, such as problem gambling and problem gaming (Bonnaire et al., 2016; Botterill et al., 2016; Edgren et al., 2016; Hardoon et al., 2004; Hing et al., 2017; Kotov et al., 2010; Michalczuk et al., 2011; Mills & Allen, 2020; Nower et al., 2004; Nower & Blaszczynski, 2017; Petry et al., 2008; Tabri et al., 2017; Vachon & Bagby, 2009). One trend which has been found through the use of SDT is the association that individuals who act with strong controlled motivations show greater problem gambling and problem gaming severity than individuals who act with autonomous ones (Allen & Anderson, 2018; Neighbors et al., 2004; Neighbors & Larimer, 2004; Rodriguez et al., 2015). Additionally, previous research has shown that need satisfaction is an antecedent of psychological well-being, while on the opposing end, need frustration has not only been associated with psychological distress but has also been linked to symptoms of multiple behavioural addictions, also including problem gaming and problem gambling (Allen & Anderson, 2018; Costa et al., 2016; Mills et al., 2018; Mills & Allen, 2020). Not to be overlooked, however, is that problematic behavioural patterns can also be influenced by various other factors, such as differences in traits, like depression, anxiety and impulsivity, as-well as psycho-social factors, such as lack of meaning in life, loneliness, poor stress-coping and more.

*The Dualistic Model of Passion and Behavioural Addictions*

Previous research on behavioural addiction has not only utilised Self Determination Theory but also the Dualistic Model of Passion and has consistently found that OP, as opposed to HP, is associated with markers of addiction and problematic consequences (Davis & Arterberry, 2019; Steers et al., 2015). For instance, HP for gaming has been linked to greater positive affect and a higher life satisfaction, while OP for gaming has been linked to negative affect, problematic behaviours, greater time spent gaming and even negative physical symptoms (Lafrenière et al., 2009; Przybylski et al., 2009; Seguin-Levesque et al., 2003; Wang et al., 2008).

*Obsessive Passions Resulting From Need Satisfaction Can Lead to Behavioural Addiction*

Situations in which global need satisfaction is low, for instance where a person feels little autonomy in their choices, limited competence in their work and marginal relatedness, have been shown to be antecedents of behavioural addiction (Allen & Anderson, 2018; Bender & Gentile, 2020; Holding et al., 2021). This is especially true, when need satisfaction within the context of a passionate activity is comparatively high, which makes the activity more likely to turn into a behavioural addiction (Mills et al., 2018; Rigby & Ryan, 2011). An example of this is video games, which allow players to make their own choices in various ways, such as designing their characters or choosing their preferred way of gameplay, thus satisfying their need for autonomy. Games are also optimally challenging, thus satisfying the need for competence and promote social relationships, thus satisfying the need for relatedness, which is, in one way, demonstrated by the massive proliferation of online multiplayer games in recent years (Ruotsalainen et al., 2022; Tomic, 2017; Zsila et al., 2022). A study with video game players, which is indicative of this, found that in-game need satisfaction was linked to higher levels of IGD, whilst global need satisfaction showed lower levels of IGD (Bender & Gentile, 2020). Furthermore, another study found that the larger the discrepancy between the context-specific and the global need satisfaction was, meaning, the more participants relied on the activity of gaming to satisfy their needs in comparison to their needs being satisfied in general life, the higher they scored on the gaming disorder scale. Hereby global need satisfaction served as a buffer to the effects of context-specific need satisfaction (Allen & Anderson, 2018). This has caused authors to postulate that addictive behaviours can develop from activity-specific need satisfaction and obsessive passion, since activity-specific need satisfaction can at times turn activities into passions (Holding et al., 2021). Consequently, a subsequent study was carried out which found that the two sources of need satisfaction, context-, or activity-specific and global, had different effects on the two types of passions. Within the framework of the activity, need satisfaction was positively linked to both passions, whereby harmonious passion showed a greater effect than obsessive passion. Furthermore, only HP was positively correlated with greater life satisfaction. The definitive role for OP seems to be played by global need satisfaction, since respondents with low global need satisfaction showed a higher probability of experiencing OP and symptoms of behavioural addiction, here in the form of gambling, even if their needs were satisfied while taking part in the activity. Thus demonstrating, in accordance with previous studies, that if need satisfaction within an activity, for example gambling or gaming, is expected to compensate for the lack of global need satisfaction in general life, this may lead the affected person to cling on to the activity. Such clinging can then develop into an obsessive passion in which the individual loses control over his behaviour toward the activity and consequently shows symptoms of behavioural addiction. The fact that global need satisfaction played no role in harmonious passion for the activity reinforces this notion (Holding et al., 2021).

**The Controversial Microtransaction: Loot Boxes**

Microtransactions, a type of monetisation in video games, are known as unrestricted purchases, usually made in-game stores, which exchange real money for additional game content or in-game virtual goods (Petrovskaya & Zendle, 2022). Loot boxes, which can also be called "prize crates", "loot crates" and various other derivative names, are a subcategory of microtransactions. They are generally defined as a consumable item of virtual nature, which, upon being redeemed, reward the player with a randomised selection of virtual items. By using the element of chance to obtain rewards, risk is introduced. Combined with a financial transaction, this is perhaps the most obvious similarity between loot boxes and traditional gambling, with some games even visually mimicking slot machines and producing similar physical responses (Bardwell, 2019; Harris et al., 2021; Larche et al., 2021). Additionally, loot boxes contain structural characteristics which appear to be psychologically similar to traditional gambling, such as the use of an intermittent reward schedule (Drummond et al., 2022; Drummond & Sauer, 2018). The items included in loot boxes can, as with all other microtransactions, range from having simply cosmetic effects ("skins"), to game-changing implications. (McCaffrey, 2022; Schwiddessen & Karius, 2018).

***The Loot Box Controversy Surrounding Problem Gambling and Problem Gaming***

Microtransactions have been proven to be very lucrative for the video game industry. For example, Activision's 2021 in-game revenue of 5.1 billion dollars made up 61% of their last year's income (Wasserman, 2022). In total, though, consumers spent an estimated $65 billion on in-game purchases in 2022. This reliance on in-game monetisation has led to the increased prevalence of the extremely profitable in-game mechanic; loot boxes, the presence of which in desktop video games has increased by 67% between 2010 and 2019 and now carries significant importance in the most established games worldwide, such as League of Legends, Call of Duty, FIFA, Battlefield and Fortnite (Zendle et al., 2020). With microtransactions constituting a big part of the gaming industry revenue and games therefore potentially being seen as a source for repeated expenditure instead of a product themselves, concern has been raised that discriminatory or exploitative monetisation techniques are being used which are not yet adequately covered by law. The main issue is that microtransaction-based revenue models are dependent on player's in-game spending which provides game designers with an incentive to concentrate on ways to advertise and encourage said spending. This kind of criticism has been expressed by multiple authors (Drummond et al., 2020; King & Delfabbro, 2018; Petrovskaya & Zendle, 2022; Zagal et al., 2013).

Different terms have been coined to describe these concepts, such as "dark patterns" and "predatory monetisation" which encompass different kinds of questionable monetisation techniques. These range from withholding the true costs of gaming activity until the players are psychologically or economically attached, to price manipulation, intrusive solicitations and in some cases even suspected use of player behaviour data to personalise the encouragement to purchase. Apart from being questionable themselves, these techniques are especially concerning when considering their potential effects on vulnerable demographics, such as children. Children are known to be more susceptible to peer pressure and the illusion that "virtual currency" can't contain real costs, as well as expressing poorer impulse control than adults (Delfabbro & King, 2020; Drummond & Sauer, 2018; Petrovskaya & Zendle, 2022; Zagal et al., 2013). This makes it especially worrying that 56% of mobile games which contain loot boxes are deemed suitable for 7 year olds (McCaffrey, 2022; Montiel et al., 2022). Such monetisation techniques, combined with the immersive nature of video games, have led to loot box research regarding the convergence of video games and gambling.

Needs, Passions and Loot Boxes – Exploring Reasons for Problem Behaviour in Relation to Loot Box Engagement***Loot Boxes, Problem Gambling and Problem Gaming***

Research on the potential parallels between problem gamblers and problem gamers, has been around since the 1990's. However, the literature on video game addiction has since then expanded into specialised areas with research on the behavioural implications of loot boxes constituting a new and contentious branch in this field (Griffiths et al., 2012; Griffiths, 1991; Macey & Hamari, 2018; McCaffrey, 2022). Two expressions which are commonly used in this branch are the aforementioned terms "problem gambling" and "problem gaming", which have both been linked to loot box engagement (Griffiths et al., 2012; Kim et al., 2017; McCaffrey, 2022; Yokomitsu et al., 2021; Zendle et al., 2020; Zendle & Cairns, 2018). With both problem behaviours being behavioural addictions there are similar symptoms to be observed, such as being preoccupied, withdrawal and loss of control (Li et al., 2019).

The academic studies on loot boxes and problem gambling arose in 2018, with Drummond and Sauer introducing potential similarities between the two. An example of this is the use of a reward structure used in many forms of gambling called a variable ratio reinforcement schedule. This reward structure manipulates the player's perception of the frequency of items gained in loot boxes. These variable ratio reinforcement schedules deliberately encourage people to acquire and frequently repeat the targeted behaviours, in the hopes of receiving their desired rewards (Ferster & Skinner, 1957; Rachlin, 1990).

The first data collected on the severity of problem gambling and loot box purchasing was done in a survey by Zendle and Cairns (2018) and demonstrated a moderate correlation between the described behaviours, which is stronger for adolescents than it is for adults (Drummond et al., 2022). This relationship has since then been replicated in different cohorts, nationalities and age groups with constant methodological improvements being made. This change in methods can also be seen as detrimental to the generalisability of the research, as will be addressed later (Spicer et al., 2022). Notably, it was particularly problem gamblers who reported reduced spending when playing games from which the loot boxes had been removed. Furthermore, less readily available purchases, for instance in card games, don't replicate the relationship between problem gambling symptoms and purchase behaviour (Drummond et al., 2022). Even though the established relationship has been replicated, solidifying a robust small to moderate relationship, the cross-sectional nature of the current data makes it hard for us to determine the direction of this relationship. This leaves us with three possibilities. A) problem gamblers buy more loot boxes, B) loot box consumers are more likely to start gambling – also known as the "gateway hypothesis" and C) there is a more complex and oscillating relationship between the two behaviours. Correlations between loot box purchasing and problem gaming, as well as correlations between problem gaming and problem gambling, are of similar magnitude, yet with more mixed results in the current literature (McCaffrey, 2022; Montiel et al., 2022; Spicer et al., 2022; Yokomitsu et al., 2021). Nonetheless, all three associations are of statistical and practical significance (Ferguson, 2009; Spicer et al., 2022). Some authors have even claimed that these relationships are often stronger than relationships between problem gambling and existing comorbidities, such as depression ($p=0.10$) and major drug problems ($r=0.12$) (Feigelman et al., 1995; Spicer et al., 2022; Zendle et al., 2019; Zendle & Cairns, 2018). However, the main findings in the literature suggest that problem gambling, problem gaming and loot box consumption are related behaviours consisting of overlapping (and diverging) drivers, paths, and potential dangers. Especially when analysing the three publications which researched all three behaviours, there is an evident emphasis on complex interaction of direct and indirect effects to be found. (Drummond et al., 2020; Drummond & Sauer, 2018; Spicer et al., 2022; Zendle et al., 2019; Zendle & Cairns, 2018). These relationships, which share bi-directional links, would not be uncommon when considering the field of behavioural addictions, since complex relationships between gambling and alternative risky behaviours are already known in the literature (Forrest & Mchale, 2018; Gupta & Derevensky, 1996; Wood et al., 2004). Due to these correlations there has been a call for the regulation of loot boxes by gamers, the media, academics and the like, which poses a problem since the "cashing out" of winnings is often considered a criterion for gambling by regulatory bodies, but which is rarely possible in the case of loot boxes (Griffiths, 2018). Nonetheless, in the UK, Brazil and Australia studies on loot boxes have been conducted with the aim of assessing potential policies towards them. The Belgian Gaming Commission also investigated loot boxes in popular games and deemed them as not giving players their "money's worth", and thus categorising them as unlicensed gambling. Similarly, the Netherlands fined EA for their use of loot boxes and removed them from their games. (Australian Government, 2019; Digital Culture Media and Sport Committee UK, 2020; Gibson et al., 2022; McCaffrey, 2022; Schwiddessen & Karius, 2018; Spicer, Fullwood, et al., 2022; Spicer, Nicklin, et al., 2022). Additionally, multiple authors have also been vocal about their wish to inform policymaking (Close et al., 2021; Drummond et al., 2020; Garea et al., 2021; Mills et al., 2021; Zendle et al., 2020; Zendle et al., 2019; Zendle & Cairns, 2018).

**Loot Boxes and Psychological Well-Being**

To date, there is little evidence concerning the actual harm of loot boxes. Though the distinct overpurchasing of loot boxes by a vulnerable group of problem gamblers suggests that they could be harmful, this does not conclusively demonstrate an association between loot boxes and harmful effects (Close et al., 2021; McCaffrey, 2022; Spicer et al., 2022). However, one recent study, using the Kessler-10 Psychological Distress Scale, was able to show that people who purchased loot boxes had surprisingly high rates of severe psychological distress, whereby the influence of age or gender was accounted for (Close et al., 2021). Interestingly, the measured rates surpassed rates of people who made other types of video game purchases, including downloadable content (DLC) and non-randomised in-game items. Surprisingly however, these rates were not exclusive to respondents with problem gambling symptoms, suggesting that those without problem gambling symptoms might be exposed to potential harm through loot box engagement as well. Finally and alarmingly, it was shown that a disproportionate amount of loot boxes were being bought by vulnerable gamers, those at a high risk of bona fide affective disorders (Drummond et al., 2022).

**Discussion**

**Loot Box Research, Self-Determination Theory and The Dualistic Model of Passion**

As established, the field of loot box research is a new and highly contentious area of study, which arose out of the public controversy surrounding loot boxes in 2017 (McCaffrey, 2022). Due to the amount of legislative, regulatory, public and academic attention which loot boxes have attracted, many studies have attempted to investigate the phenomena of loot boxes, despite their short existence. A downside to the emotionally charged nature of the loot box literature is that multiple authors are reaching premature and probably biased conclusions which don't accurately represent the data. What some studies have shown, however, is that loot boxes are a new and, for some players, exciting, even physiologically arousing, game-mechanic on which, in some cases, people spend large amounts of real-world money (Close et al., 2021; Larche et al., 2021). Apart from physiological excitement and reward reinforcement schedules, loot boxes also share some aesthetic similarities with traditional gambling, mainly slot machines, with the use of exciting images and sounds (Drummond et al., 2020; Drummond & Sauer, 2018). More importantly, however, the correlation between loot box engagement and problem gambling or problem gaming is a robust finding, which has been replicated in several studies and could predict a significant pattern of behaviour (McCaffrey, 2022; Spicer et al., 2022). This suggests that problem gambling, problem gaming and loot box engagement are related behaviours which overlap in various ways and are in complex interaction with each other (Spicer et al., 2022). These correlations have been indirectly linked to psychological harm, though these interpretations have been addressed with caution (Drummond et al., 2020; Li et al., 2019; Spicer et al., 2022). To date, however, there has not been an actual case of problem gaming or problem gambling with respect to loot boxes, as all of the empirical evidence has been found through the use of self-reported surveys. Consequently, the existing evidence is not yet indicative of loot boxes being responsible for widespread and consistent harm to players (Close et al., 2021; McCaffrey, 2022; Spicer et al., 2022).

*Loot Box Research Methodology*

Picking up the notion of the convergence of video gaming and gambling, which was introduced in the 1990s, the current literature has tried to apply this same framework to the study of loot boxes (Gibson et al., 2022; Griffiths, 1991, 2018). This has been inherently difficult due to the extremely diverse nature of loot box forms, availability, value to players, engagement prevalence and conscious motivations for engagement. Finding a way to study unique aspects of game mechanics and the unique decisions which loot box users face is one of the main difficulties with which the loot box literature has been confronted. The cross-sectional data from the self-reported surveys, which are vulnerable to selection bias, only allows for correlational interpretations and isn't widely generalisable due to methodological issues. Systematic reviews of the literature have advised caution in making inferences based on the current data, due to implications that the methodological quality of most studies is fairly low. No studies have yet been rated as being of the highest quality (Yokomitsu et al., 2021). Furthermore, most studies assume that loot boxes can be studied using methods developed for studying traditional gambling behaviour, without making any special modifications to capture the unique features of the loot box mechanic (McCaffrey, 2022).

***Self-Determination Theory in Relation to Loot Box Engagement, Problem Gambling and Problem Gaming***

An area which has been undermined in the context of loot box engagement is the influence of needs and motivations which have both been linked to behavioural addiction (Holding et al., 2021; Mills et al., 2021; Mills et al., 2018; Mills & Allen, 2020). In a study by Mills et al. (2021) the conceptual framework of Self-Determination Theory was used to study behavioural addiction in the context of problem gamblers. The study highlighted the influence that general motivations and need frustration play in the development of addictive behaviour. Hereby a greater frustration of psychological needs was associated with a higher risk for problem gambling. Interestingly, need frustration has also been shown to be an antecedent to problem gaming (Allen & Anderson, 2018). Moreover, the positive effect which amotivation and controlled motivations have on problem gambling and psychological distress, seem to be explained by an increase in need frustration. If this is cohesive with SDT's fundamental assumption that general motivations can lead to both adaptive or maladaptive outcomes through experience of need satisfaction or need frustration, this could define need frustration as a mediator for predicting problem gambling and psychological distress (Mills et al., 2021; Ryan & Deci, 2017). Stronger controlled motivations lead to higher need frustration which in turn may lead to higher levels of problem gambling or problem gaming, as well as psychological distress.

Loot boxes in video games are often used as a way to progress faster or pose a competitive threat to other players. However, these loot boxes often have to be bought using real-world money and, especially with games which were already bought at a premium price level, it is not always easy

Needs, Passions and Loot Boxes – Exploring Reasons for Problem Behaviour in Relation to Loot Box Engagementto understand why players would spend such high amounts of money on them. A quick reminder at this point that loot box sale estimations for 2022 exceed $18 billion (Clement, 2021a) which would account for almost a third of last year's worldwide in-game consumer spending (Clement, 2021b). With the application of the SDT framework and its theory of psychological needs, one could argue that in accordance with the existing relationship between loot box engagement and problem gamblers or problem gamers, players which show high loot box engagement and, therefore, possibly symptoms of problem behaviour, are doing so due to need frustration. The competence need could hereby portray the game's difficulty or the players' ability, or inability, in the game. For instance, if a player has reached a certain level in a multiplayer game and is constantly matched up with other players who are better than he is, this may frustrate his basic need for competence, making him feel inadequate at the game. This may also influence his basic need for autonomy by forcing him to play a certain way to pose a competitive threat, constraining his freedom of playstyle. This need frustration could lead players to use controlled motivations to find a way to re-satisfy their needs in the game (Mills & Allen, 2020; Ryan & Deci, 2017; Vansteenkiste et al., 2020; Vansteenkiste & Ryan, 2013; Vuorinen et al., 2022). One easily accessible way to do this would be to buy loot boxes, the contents of which will improve their chances of winning or being better at the game, since loot boxes often contain content which has an influence on gameplay, in this example, providing a competitive edge (King et al., 2019; King & Delfabbro, 2018; Tomić, 2019; von Meduna et al., 2020). Since need frustration acted as a mediator in the study by Mills et al. (2021) this would be equally applicable to the proposed scenario, with the in-game need frustration leading to psychological distress as well as problem gambling or problem gaming, in the form of opening loot boxes in the hope of receiving better rewards or playing the game incessantly. This would also align with the notion that individuals who act with controlled motivations have a need to demonstrate their self-worth, here, in form of their status in-game (Hodgins & Knee, 2002; Kernis, 2003; Lewis & Neighbors, 2005).

***Global vs Context-Specific Need Satisfaction and Passions in Relation to Loot Boxes***

Expanding upon the opposite side of need frustration are the two subcategories of need satisfaction; global need satisfaction and context-specific need satisfaction. Global need satisfaction represents needs in general life, whereby context-, or activity-specific need satisfaction represents needs satisfied through a specific activity (Allen & Anderson, 2018; Holding et al., 2021; Lalande et al., 2017). In the previously mentioned study by Allen & Anderson (2018), respondents who relied more on context-specific need satisfaction by way of gaming, as opposed to need satisfaction in general, scored higher on the Internet Gaming Disorder scale. These findings are in accordance with other past research, which has shown that if need satisfaction within an activity is comparatively high compared to need satisfaction outside the activity, the activity is more likely to turn into a behavioural addiction (Lalande et al., 2017; Mills et al., 2018; Rigby & Ryan, 2011). As previously discussed, past research has shown that global need satisfaction is negatively associated with OP, meaning individuals whose needs aren't met outside a specific activity score higher in levels of OP. This suggests that people whose general needs aren't met could turn to an activity which, in compensating fashion, satisfies their needs and subsequently develops into an obsessive passion (Holding et al., 2021; Seguin-Levesque et al., 2003; Vallerand, 2015; Wang et al., 2008).

      This would allow for a different explanation of the relationship between problem gamblers and problem gamers, by including the notion of obsessive passions. In this case the act of playing video games would serve as their context-specific need satisfaction. Playing games and, for instance in a competitive setting, being good at the game would satisfy the needs which they lack in general life. This would consequently turn gaming into an obsessive passion, an activity from which they derive their needs, with which they also feel an uncontrollable urge to interact, even if it is in conflict with other goals and responsibilities (Holding et al., 2021; Mageau & Vallerand, 2007; Seguin-Levesque et al., 2003). As seen in previous research, people who derive more need satisfaction from a passionate activity than from life in general are prone to have that activity turn into an obsessive passion and consequently show symptoms of behavioural addiction, such as problem gambling or problem gaming (Mills et al., 2018; Rigby & Ryan, 2011). To have their needs satisfied constantly, therefore constantly being good at the game, players could use loot boxes as a way of preserving their status in-game. Due to the continued release of new and better items over the course of a game's life cycle this could ultimately force these players with obsessive passions into buying more loot boxes. This is especially true for games which have shown to be almost impossible to complete without investing ludicrous amounts of either time or money in (King et al., 2019; Petrovskaya & Zendle, 2022). This group of players with obsessive passions and problem behaviour could in fact account for the majority of loot box consumers, since a third of high-level spenders are problem gamblers, with problem gamers being of a similar magnitude. These high-level spenders account for half of total loot box revenue (Close et al., 2021; Garea et al., 2021).

**Limitations in the literature**

Even though there have been repeated findings of a relationship between problem gambling and loot box consumption, there are multiple negatives to be mentioned, especially methodological ones, which prevent the current literature on loot box engagement from being generalisable. One of these is reliable sampling, with almost all studies on loot box engagement using cross-sectional data from self-reported surveys. One example of this would be multiple studies, claiming to produce research on games and gambling, using data which was gathered by posting survey links on Reddit (Zendle et al., 2019; Zendle & Cairns, 2018). This provides potential for self-selection bias, due to the animosity surrounding loot boxes in the public debate. An illustration of this might be the most downvoted comment on the Reddit website, which is that of an EA employee defending the company's use of loot boxes in their game "Battlefront 2" (Koepp, 2021). Even though the limitations of utilising platforms such as Reddit for this research have been acknowledged, they continue to be used in new studies (McCaffrey, 2022).

Another problem concerning the current literature is the suboptimal choice in demographics, which impairs the representativeness and generalisability of the studies. Only about a third of all studies include adolescents aged 12-17, with most using data from adult surveyees and none involving younger children (DeCamp, 2021; Kristiansen & Severin, 2020; Rockloff et al., 2021; Zendle et al., 2019). This is not surprising considering the difficulty of conducting research with children, but it is still regrettable, as children represent the demographic with the highest vulnerability towards loot boxes. This is especially alarming when considering that a survey by the UK Gambling Commission showed that 31% of children had already opened loot boxes and that 93% of games that feature loot boxes are marked suitable for ages 12 and up. (Advisory Board for Safer Gambling UK, 2021; McCaffrey, 2022; Montiel et al., 2022). We can conclude that only about a fifth of empirical data can be judged as using representative samples with a good prospect for generalisability (McCaffrey, 2022). Furthermore, the actual prevalence of loot box engagement is unclear, due to the varied reports in the current literature, which most likely differ due to the use of different sampling methods. These reports range from 3.5% of respondents merely opening loot boxes to 60.3% of respondents actually going to the length of paying for loot boxes (Brooks & Clark, 2019; Ide et al., 2021).

Moreover, the screening and measurement tools used to assess the problem behaviour are not ideal. One problem is the failure to isolate consumer spending on the uniquely gambling-like features of loot boxes as opposed to high spending by consumers in general. This leads to ambiguity concerning general spending, as players could be high or low spenders, who spend large or small amounts on loot boxes. Additionally, as previously mentioned, the PSGI is often used to measure symptoms of gambling disorder in loot box research. This could be problematic, as it has been proven to be difficult to distinguish between low-risk and medium-risk gamblers, nor is it an optimal tool for studying loot boxes (Macey & Hamari, 2018; McCaffrey, 2022).

**Future Research**

With loot box research effectively originating from a public controversy it is not hard to grasp the fact that most of the research seems to be constrained by the emotionally charged agenda for loot box regulation. This bias has led to insufficient transparency in the field, a point that should be emphasised for future research. Loot box studies should define more clearly what they do and do not study and what the potential limitations are. This includes aiming for a more accurate characterisation of key demographics, which in turn constitutes conducting research on the most vulnerable of loot box demographics, children. To gain a clearer idea of causation it is also essential to generate and investigate longitudinal data to examine what the relationship between loot box engagement and problem behaviour looks like over time. Additionally, future studies on loot boxes should enforce conditions which limit the chance for sample selection bias and thus eliminate the influence of respondents with strong pre-existing opinions on the matter.

Furthermore, instead of hastily trying to apply former methods for measuring problem behaviour to loot boxes, different and innovative methods should be considered to reflect the novel assets which video games, microtransactions and loot boxes possess. These should also account for the player's unique choices when engaging with loot boxes and the underlying motivations that encourage him to do so. Hereby an expansion upon the study by Holding et al. (2021) which would further analyse the relation between need satisfaction, obsessive passion and loot box engagement could provide insight into possible reasons for problem behaviour in relation to loot boxes and in turn give us a better understanding of these interactions and how to minimise their potential harm on players (Garea et al., 2021; Gibson et al., 2022; Macey & Hamari, 2018; McCaffrey, 2022; Montiel et al., 2022; Petrovskaya et al., 2022; Schwiddessen & Karius, 2018; Spicer et al., 2022; Xiao & Henderson, 2021; Yokomitsu et al., 2021).

**Conclusion**

The goal of this thesis was to demonstrate the relationship between loot boxes and problem gamblers or problem gamers and, due to the inconclusiveness of the current literature, postulate how this relationship could be better explained. In doing so, this thesis posited the potential influence of need satisfaction and need frustration and the resulting development of harmonious or obsessive passions on behavioural addictions, such as problem gambling and problem gaming in relation to loot box engagement. Due to the cross-sectional nature of current loot box data, conclusions about causality are limited. Future research using longitudinal data and the incorporation of the Self-Determination Theory framework, combined with the notions of the Dualistic Model of Passion, could provide a better understanding of this complex relationship and what measures could be taken to minimise potential harm, especially in form of behavioural addiction, being done to players.

Needs, Passions and Loot Boxes – Exploring Reasons for Problem Behaviour in Relation to Loot Box Engagement